\documentclass[pre,superscriptaddress,twocolumn,a4paper,showpacs]{revtex4-1}
\usepackage{graphics}
\usepackage{amssymb}
\usepackage{amsmath}
\usepackage{wasysym}
\usepackage{latexsym}
\usepackage{graphicx}
\usepackage{epsfig}
\usepackage{subfigure}
\usepackage{float}
\usepackage{color}
\usepackage{lineno}
\usepackage{verbatim}
\usepackage{enumitem}

\begin{document}

\title{A soluble model for synchronized rhythmic activity in ant colonies}

\author{Pedro M.M. da Silveira}
\affiliation{Instituto de F\'{\i}sica de S\~ao Carlos,
  Universidade de S\~ao Paulo,
  Caixa Postal 369, 13560-970 S\~ao Carlos, S\~ao Paulo, Brazil} 

\author{Jos\'e F.  Fontanari}
\affiliation{Instituto de F\'{\i}sica de S\~ao Carlos,
  Universidade de S\~ao Paulo,
  Caixa Postal 369, 13560-970 S\~ao Carlos, S\~ao Paulo, Brazil}


\begin{abstract}
Synchronization is one of the most striking instances of collective behavior, occurring in many natural phenomena. For example, in some ant species, ants are inactive within the nest most of the time, but their bursts of activity are highly synchronized and involve the entire nest population. Here we revisit a simulation model that generates this synchronized rhythmic activity through autocatalytic behavior, i.e., active ants can activate inactive ants, followed by a period of rest. We derive a set of delay differential equations that provide an accurate description of the short- and medium-time simulations for large ant colonies.  Analysis of the fixed-point solutions, complemented by numerical integration of the equations,  indicates the existence of  stable limit-cycle solutions  when the rest period is greater than a threshold and the event of spontaneous activation of inactive ants is very unlikely, so that most of the arousal of ants is done by active ants. Furthermore, we argue that the persistent oscillations observed in the simulations for colonies of finite size are due to resonant amplification of demographic noise.
\end{abstract}

\maketitle

%
\section{Introduction} \label{sec:intro}
 %
 
Perhaps the most remarkable example of self-organization in nature is the endogenous (i.e., not driven by an external signal) synchronization observed in abiotic oscillatory chemical reactions \cite{Nicolis_1977} and in population dynamics laboratory experiments \cite{Nicholson_1954,May_1975}, where reactants and individuals have no intrinsic activity rhythm.
 Much of our understanding of synchronization, however, comes from the study of coupled oscillators, where each oscillator has its own intrinsic natural frequency \cite{Strogatz_1993,Strogatz_2000}.

A well known and documented case of endogenous synchronization, at least within the entomological community, is the activity bursts of individuals in colonies of some ant species \cite{Franks_1990,Cole_1991,Doering_2022}. For example, ants of the species Leptothorax acervorums are inactive within the nest for about 72\% of the time, but their bursts of activity are highly synchronized, occurring about three or four times per hour and involving the entire nest population \cite{Franks_1990}.  We  use the terms nest and colony interchangeably because all individuals in a colony share food in a nest, but only a few go out to collect it.  The care of the brood is the main task of the worker ants within the nest.

There have been two major attempts to provide a mathematical model for the emergence of these short-term activity cycles. The first assumes that they are caused by the need for energy (food) to meet the needs of the nest \cite{Hemerik_1990}. In particular, in this approach the rate of increase of the energy level $E$  is assumed to be proportional to the number of active ants $A$  inside the nest, but   the rate of increase of $A$ is determined by an ad hoc nonlinear function of $A$ and $E$, specifically designed to produce stable limit cycles  and thus of difficult biological justification \cite{Hemerik_1990}.  We will not consider this model here. The second approach, which we call the autocatalytic ant colony model \cite{Goss_1988},  builds on the well-established experimental fact that ants can be stimulated by nest mates to become active after a refractory rest period  \cite{Franks_1990,Cole_1991,Doering_2022}.  Numerical analyses of the autocatalytic ant colony model using Monte Carlo simulations \cite{Goss_1988} and  of a simpler variant using process algebra \cite{Tofts_1992} indicate that the model seems to be  capable of generating synchronized bursts of activity. 

Here we derive a set of delay differential equations that accurately describe the Monte Carlo simulations of the autocatalytic ant colony model in the limit of very large colony size but not too long time. Analysis of the stability of the fixed-point solutions shows that stable limit-cycle solutions appear when the rest period is greater than a threshold and the event of spontaneous activation of inactive ants is very unlikely, so that most of the arousal of ants is done by active ants. We find that the range of model parameters for which limit-cycle solutions exist is very limited, and thus the synchronized bursts of activity observed in Monte Carlo simulations of colonies of finite size \cite{Goss_1988} is a resilient  effect of demographic noise  known as  coherence resonance \cite{McKane_2005,Kuske_2007}. These stochastic fluctuations prevent the decrease in amplitude of the damped oscillations that would eventually bring the dynamics to the deterministic equilibrium.

The rest of the paper is organized as follows.   In Section \ref{sec:model}, we introduce the autocatalytic ant colony model and describe how asynchronous Monte Carlo simulations of this model are implemented. Section \ref{sec:DE} presents our main contribution, namely, the derivation of a set of mean-field delay differential equations that describe the autocatalytic ant colony model in the limit of infinite colony sizes. Comparison between the results of the numerical integration of these equations and the Monte Carlos simulations reveals very robust finite-size effects for long times due to resonant amplification of demographic noise. The  technical  Section \ref{sec:fp} presents the stability analysis of the fixed point solutions of the mean field equations, where we obtain the regions in the parameter space where the periodic solutions appear. In Section  \ref{sec:per}, we study the period of the oscillatory solutions with vanishingly small amplitudes. Finally, in Section \ref{sec:conc} we summarize our results and give some concluding remarks.

%
\section{Autocatalytic ant colony model}\label{sec:model} 
%
Following the seminal work that introduced the autocatalytic ant colony model \cite{Goss_1988}, we consider a colony of $N$ ants, where each ant can be in one of three possible states: inactive, activatable inactive, and active. Once an ant becomes inactive, it remains in that state for a fixed period $\tau$, which we call the rest period. When the rest period is over, the inactive ant becomes an activatable inactive ant. An activatable inactive ant can either become active spontaneously with a rate of $\alpha$ or, more importantly, can be activated by an active ant. Hence the positive feedback or autocatalytic property of the model. The  effectiveness of  the autocatalytic activation is determined by the parameter $\beta$.  An active ant becomes inactive at a rate  of $\mu$. The use of the same rest period for all ants is a simplification of the model, as it has been observed that the duration of rest periods varies between species and can be highly variable within the same species \cite{Doering_2022}.

The asynchronous Monte Carlo simulation of the ant colony from  time $t$ to  $t+\delta t$ is as follows. At time $t$, 
we pick an ant at random, say ant $i$, and check its state.  Suppose ant $i$ is activatable inactive at time $t$. Then  it becomes active at time $t+ \delta t$ with probability $\alpha$. Next, assume that  ant $i$ is active. Then there are two actions that take place sequentially. First,  ant $i$ randomly chooses another ant in the colony, say ant $j$: if ant $j$ is activatable inactive then  ant $j$ becomes active at time $t+\delta t$ with probability $\beta$. Second,   with probability $\mu$, ant $i$ becomes inactive at time $t+ \delta t$.   Finally, suppose that ant $i$ is inactive.  If it has been at rest for a time less than $\tau$, it remains inactive at time $t+ \delta t$, but if it has been at rest for a time equal to $\tau$, it becomes activatable inactive at time $t+ \delta t$. At this point we can see that the time step $\delta t$ must be a divisor of the rest period $\tau$. In fact, regardless of whether it is selected for update or not, at each time step we check each inactive ant and move it to the activatable inactive state if it has been at rest for a time equal to $\tau$.
As usual in such an asynchronous update scheme, we choose the time increment as $\delta t = 1/N$. Thus, the product $N \tau$ must be an integer. In Section \ref{sec:DE}  we will give an alternative justification for this choice of the time step $\delta t$.

Monte Carlo simulations for small colonies produced regular bursts of synchronized activity (see Fig.\ \ref{fig:1}), but if the rest period $\tau$ was too short relative to the activity period $1/\mu$, or if the activation coefficient $\beta$ was too low relative to the spontaneous activation rate $\alpha$, the simulations produced unsynchronized activity \cite{Goss_1988}.  As we will see in Section \ref{sec:fp}, the unsynchronized activity observed in the simulations corresponds to the fixed-point solutions of the mean-field equations.

%
\section{Mean-field equations}\label{sec:DE} 
%

We begin our derivation of the equations governing the dynamics of the autocatalytic ant colony model in the $N \to \infty$ limit by specifying the ant state transition rules in a more formal way.   Let us first introduce some notation: $a_i(t)$, $b_i(t)$, and $s_i(t)$ are the probabilities that ant $i$ is active, activatable inactive, and inactive, respectively, at time $t$. Obviously, $a_i(t) + b_i(t) + s_i(t) = 1$.

The probability $a_i(t+\delta t)$ that ant $i$ is active at time $t+\delta t$ is given by the sum of the probabilities of the next four (exclusive) events.
 \begin{enumerate}[label=(\alph*)]
\item Ant $i$ is selected for update and  is active at time $t$ and remains active. The probability of this event is $1/N \times a_i(t) \times (1-\mu)$.
\item Ant $i$ is selected  for update and is activatable inactive at time $t$ and becomes spontaneously active. The probability of this event is $1/N \times b_i(t) \times \alpha$.
\item Ant $i$ is active at time $t$ and any ant except ant $i$ is selected for update. The probability of this event is $a_i(t) \times (N-1)/N$.
\item Ant $i$  is activatable inactive at time $t$ and an active ant is selected for update, which then activates ant $i$. The probability of this event is $ b_i(t)  \times A(t)/N \times 1/(N-1) \times \beta $, where $A(t)$ is the number of active ants at time $t$.
 \end{enumerate}
Adding the probabilities of these events yields
\begin{equation}\label{ai0}
 a_i(t+ \delta t) = a_i(t)- \mu \frac{1}{N} a_i(t)  + 
\alpha   \frac{1}{N} b_i(t) +
\beta \frac{1}{N} b_i(t)   \frac{A(t)}{N-1}  .
\end{equation}

Now we consider the  more difficult derivation of the probability $b_i(t+\delta t)$ that ant $i$ is activatable inactive at time $t+\delta t$. This probability  is given by the sum of the probabilities of the next five (exclusive) events.
 \begin{enumerate}[label=(\alph*)]
\item Ant $i$ is selected for update and  is activatable inactive at time $t$ and  does not become active spontaneously. The probability of this event is $1/N \times b_i(t) \times (1-\alpha)$.

\item Ant $i$ is activatable inactive at time $t$ and any inactive ant or any activatable inactive ant except ant $i$ is selected for update.    The probability of this event is $b_i(t) \times [S(t) + B(t) -1]/N$, where $S(t)$ and $B(t)$ are the number of inactive and activatable inactive ants, respectively, at time $t$.

\item Ant $i$ is activatable inactive at time $t$ and an active ant is selected for update, but this ant does not select  ant $i$ for activation.    The probability of this event is $b_i(t) \times A(t)/N \times (N-2)/(N-1)$. 

\item Ant $i$ is activatable inactive at time $t$ and an active ant is selected for update and  this ant selects  ant $i$ for activation, but the activation fails.    The probability of this event is $b_i(t) \times A(t)/N \times 1/(N-1) \times  (1-\beta)$. 

\item Ant $i$ was active  and was selected for update at time $t-\tau$ when it became inactive. The probability of this event is $a_i(t-\tau) \times 1/N \times \mu$.
\end{enumerate}
Adding all these probabilities yields
\begin{equation}\label{bi0}
b_i(t + \delta t) = b_i(t) -  \frac{1}{N} b_i(t) \left ( \alpha + \beta \frac{A(t)}{N-1} \right )  + \mu \frac{1}{N} a_i(t-\tau),
\end{equation}
where we have used $A(t) + B(t) + S(t) = N$. The equation for $s_i(t + \delta t)$ can now be obtained using normalization.

To finish setting up the model, we note that $a_i(t+\delta t) -a_i(t)$ and $b_i(t+\delta t) - b_i(t)$ must be proportional to $\delta t$, which is the case if we set $\delta t = 1/N$. Finally, taking the limit $N \to \infty$, we find 
\begin{equation}\label{aif}
\frac{da_i}{dt}= - \mu a_i(t)  +  \alpha  b_i(t) +  \beta b_i(t)  \rho_a ,
\end{equation}
\begin{equation}\label{bif}
\frac{d b_i}{dt} =  -  \alpha b_i(t)   - \beta  b_i(t)  \rho_a  + \mu  a_i(t-\tau),
\end{equation}
\begin{equation}\label{sif}
\frac{d s_i}{dt} = \mu  \left [ a_i(t) - a_i(t-\tau) \right ]
\end{equation}
with the normalization $a_i(t) + b_i(t) + s_i(t) = 1$ for $i=1, \ldots ,N$ and 
\begin{equation}\label{rho}
\rho_a (t) = \lim_{N \to \infty} \frac{A(t)}{N}.
\end{equation}
Note that the coupling between different ants is due to the term $\rho_a (t)$. This is clear if we rewrite $A(t)$ using an indicator function, viz.,  $A(t) = \sum_{j=1}^N \chi_j$ where $\chi_j =1 $ if ant $j$ is active and $\chi_j =0 $ otherwise.

Assuming that individual ants are indistinguishable, we can write $a_i(t) = a(t)$, $b_i(t) = b(t)$ and $s_i(t)= s(t)$ for $i=1, \ldots N$, so that $a(t)$ can be interpreted as  the expected proportion of active ants at time $t$. If we equate this expected value with the fraction of active ants at time $t$, i.e., with the ratio $\rho_a $, we obtain a closed set of equations for the colony dynamics. This is the usual mean-field approximation of statistical physics (see, e.g., \cite{Huang_1963}), which in our case becomes  exact in the limit $N \to \infty$ due to the law of large numbers \cite{Feller_1968}.  A similar interpretation holds for the fraction of inactive $s(t)$ and activatable inactive $b(t)$ ants. Henceforth, we will omit the qualifier mean or expected when referring to the fraction of ants in different states, since we will assume that $N$ is sufficiently large for the mean-field approximation to be valid.  Rewriting the equations (\ref{aif}), (\ref{bif}) and (\ref{sif}) using this approximation yields
\begin{eqnarray}
\frac{da}{dt} & = & - \mu a(t)  + b(t) \left [  \alpha  +  \beta a(t) \right ] \label{a_det} \\
 \frac{d b}{dt} & = &  -  b(t)  \left [ \alpha  + \beta a(t) \right ]  + \mu  a(t-\tau) \label{b_det} \\
 \frac{d s}{dt}  & =&  \mu  \left [ a (t) - a (t-\tau) \right ] \label{s_det}
\end{eqnarray}
with $a(t) + b(t) + s(t) = 1$.  Although $s(t)$ does not appear in the equations for $a(t)$ and $b(t)$, Eq. (\ref{s_det}) will be crucial in determining the fixed-point  solutions, as we will see next.  We can set $\beta =1$ without loss of generality by measuring all other parameters in units of $\beta$ and rescaling the time accordingly.  However, this is only done in the numerical analysis.

In this paper we use the initial conditions $a(t)=0$ for $t \in [-\tau,0)$, $a(0)=1$, $b(0)=0$ and $s(0)=0$, but this choice does not affect our results since the fixed point solutions of the mean-field equations do not depend on the initial conditions. In particular, for this choice we get $a(t) = \exp(-\mu t)$, $b(t) = 0$ and $s(t) = 1 - \exp(-\mu t)$ for $t \in [0, \tau]$.

It is interesting to note that  the time-delay mean field equations (\ref{a_det}), (\ref{b_det}),  and (\ref{s_det}) are  very similar to  those used in SIRS models of epidemiology, where individuals transition between three different states (viz., susceptible, infected, and recovered) and,  after infection, individuals become immune for a fixed period of time and then  transition deterministically to the susceptible state \cite{Hethcote_1981,Sebastian_2011}.   This similarity is understandable because in the SIRS model infected individuals can spread the disease by contact, which is analogous to the spread of activity in the autocatalytic ant model.  A difference between the two models is that ants can spontaneously activate at the rate $\alpha$, so activity, unlike disease, can reappear even if all the ants eventually become inactive.

\begin{figure}[t] 
\centering
 \includegraphics[width=1\columnwidth]{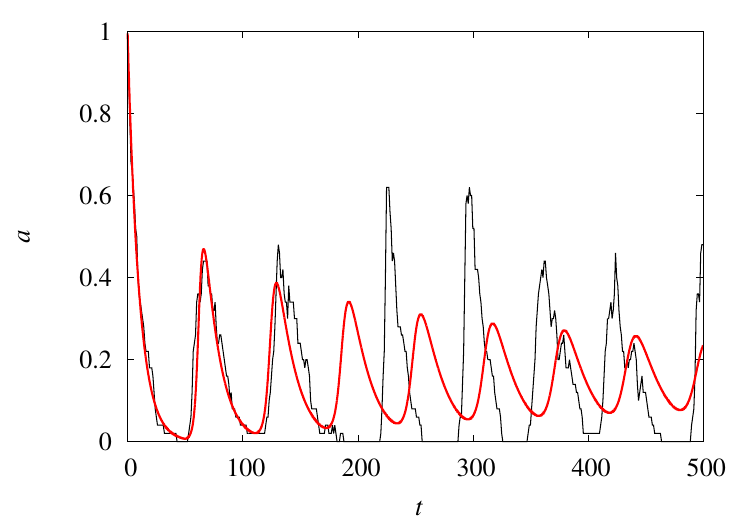}  
\caption{Time evolution of the  fraction of active ants $a(t)$  for a population of size  $N=50$ and  rest period $\tau=49$. The black jagged thin curve is the result of a single run of the Monte Carlo simulation and the  red smooth thick curve is the numerical solution of the mean-field equations. The other parameters are  $\mu = 1/10$, $\alpha = 1/500$ and $\beta=1$.   The initial conditions are $a(t)=0$ for $t \in [-\tau,0)$, $a(0)=1$, $b(0)=0$ and $s(0)=0$.  
 }  
\label{fig:1}  
\end{figure}

Before analyzing the fixed-point solutions, it is instructive to compare the numerical solution of the mean-field equations with the Monte Carlo simulations.  Figure \ref{fig:1} shows the time evolution of the fraction of active ants for a colony of size $N=50$ and rest period $\tau=49$.  This is very close  to the rest period $\tau=50$ used in the original simulation of the ant colony model \cite{Goss_1988}. In fact, there are no noticeable differences between the simulation results or the numerical solutions for these two choices of rest period. The reason why we do not use $\tau=50$ will become clear later when we discuss the stability of the fixed-point solutions.
 The bursts of activity occurring at intervals roughly equal to the rest period $\tau$ were interpreted as evidence of synchronized periodic activity  \cite{Goss_1988}. This pattern, and in particular the amplitude of the activity peaks, remains unchanged as we follow the Monte Carlo dynamics further in time.  This result is in stark contrast to the numerical solution of the mean-field equations, which show a damped oscillation that eventually leads to a fixed point. This discrepancy is of course due to the small population size used in the simulations.

\begin{figure}[h] 
\centering
 \includegraphics[width=1\columnwidth]{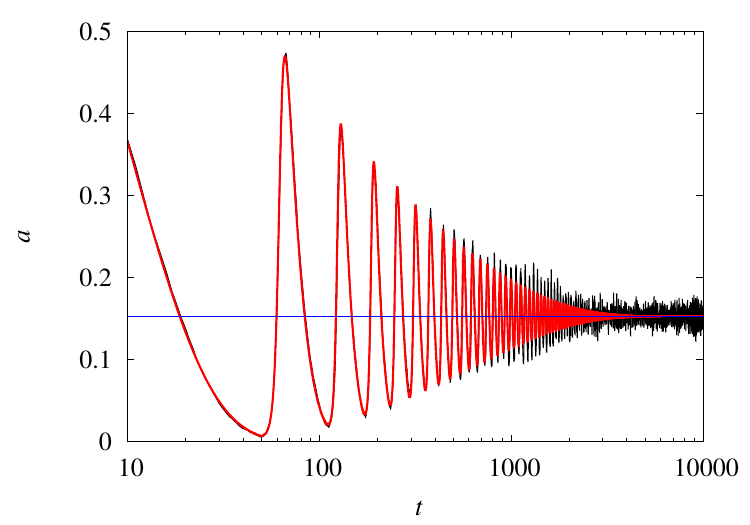}  
\caption{Time evolution of the  fraction of active ants $a(t)$    for a population of size  $N=5000$ and rest period $\tau=49$. The black jagged thin curve is the result of a single run of the Monte Carlo simulation and the red smooth thick curve is the numerical solution of the mean-field equations. The other parameters are  $\mu = 1/10$, $\alpha = 1/500$ and $\beta=1$.    The fixed point is $a^* \approx  0.153$ (thin horizontal line), $b^* \approx 0.099$ and $s^* \approx  0.748$. The initial conditions are $a(t)=0$ for $t \in [-\tau,0)$, $a(0)=1$, $b(0)=0$ and $s(0)=0$.  
 }  
\label{fig:2}  
\end{figure}

In fact, Fig.\ \ref{fig:2} shows that the results of a single run of the Monte Carlo simulation with $N=5000$ ants agree very well with the predictions of the mean-field equations except in the asymptotic time limit, when  instead of converging to a stable equilibrium, the simulation keeps oscillating forever around the fixed point.
Although the amplitudes of the oscillations decreases with increasing $N$ (compare Fig.\ \ref{fig:1} for $N=50$ with Fig.\ \ref{fig:2} for $N=5000$), they are not of the order of $1/\sqrt{N}$ as one would usually expect.  In fact, the persistent oscillations observed in the simulations for finite populations, even when the deterministic solutions show damped oscillations, are due to a mechanism called coherence resonance or  autonomous stochastic resonance: the internal demographic noise, by exciting all frequencies, automatically resonates the system leading to large oscillations in the population densities \cite{McKane_2005,Kuske_2007}.
 As shown in  Fig.\  \ref{fig:2}, the path of a Monte Carlo run  with a large population follows the damped trajectory of the corresponding deterministic model for some initial time, after which the stochastic path remains oscillatory.  These oscillations have a narrow frequency distribution and stochastically varying amplitude \cite{McKane_2005}.

The results presented in Figs.\ \ref{fig:1} and \ref{fig:2} raise the question of whether the autocatalytic ant colony model exhibits truly periodic solutions in the limit of infinitely large populations. The answer to this question requires a study of the fixed point solutions of equations (\ref{a_det})-(\ref{s_det}).

\section{Fixed point analysis}\label{sec:fp}

The fixed point or equilibrium solutions  $a^* = a(t)$, $ b^* = b(t)$ and $s^* = s(t)$   are obtained by setting the time derivatives to zero in the mean-field  equations  (\ref{a_det}),  (\ref{b_det}) and  (\ref{s_det}). The difficulty here is that this procedure yields only one equation, viz.,
\begin{equation}\label{fp_b}
b^* = \frac{\mu a^*}{ \alpha  +  \beta  a^*},
\end{equation}
and  so we need one more equation to determine the equilibrium solution unambiguously, since $a^* + b^* + s^* =1$. The missing equation is found by integrating Eq. (\ref{s_det}) from $0$ to $n \tau$ where $n$ is an integer, 
\begin{equation}
\int_0^{n \tau} \frac{d s}{dt}  dt = \mu  \left [ \int_0^{n\tau} a (t) dt - \int_0^{n\tau} a (t-\tau) dt\right ] ,
\end{equation}
which, after some elementary manipulations, can be  rewritten as 
\begin{equation}
s(n\tau) - s(0)  = \mu  \left [   \int_{n \tau - \tau}^{n \tau} a (t) dt - \int_{-\tau}^{0}  a (t) dt \right ] .
\end{equation}
Now, taking the limit $n \to \infty$ and assuming that the dynamics reach the equilibrium solution, we get  
\begin{equation}
s^* - s(0)  = \mu  \left [   \tau a^* - \int_{-\tau}^{0}  a (t) dt \right ] .
\end{equation}
We recall that for $t \in [-\tau,0)$ the fraction of active ants $a(t)$ is  given by the initial conditions, which in turn determines the fraction of inactive ants at $t=0$, viz., $s(0) = \mu \int_{-\tau}^{0}  a (t) dt $. Therefore,
\begin{equation}\label{fp_s}
s^*  = \mu  \tau a^* 
\end{equation}
regardless of the initial conditions. The intuition for this result is clear: at equilibrium,  individuals become inactive at the rate $\mu a^*$ and remain inactive for the period $\tau$.
Using equations  (\ref{fp_b}) and (\ref{fp_s}) together with the normalization condition gives  the  equation for the fraction of active ants at equilibrium,
\begin{equation}\label{fp_a}
\beta (1+\mu \tau) (a^*)^2 
+  \left [ \alpha (1+\mu \tau) + \mu -\beta  \right ] a^*- \alpha = 0.
\end{equation}
This quadratic equation always has real roots:   one negative and the other positive less than $1$, so the fixed points  obtained with the equations (\ref{fp_b}), (\ref{fp_s}) and (\ref{fp_a}) are always valid equilibrium solutions of the mean-field equations, i.e., $a^*$, $b^*$ and $s^*$  are positive and less than $1$  for all parameter settings.  As a good approximation, we can write $a^* \approx 1/\mu \tau $,
$b^* \approx 1/\alpha \tau$  and $s^* \approx 1 - (\mu + \alpha)/\mu \alpha \tau$, which holds for $\mu \tau \gg 1$.

To study the stability of the fixed point solutions \cite{Murray_2007}, we must first linearize equations (\ref{a_det}) and (\ref{s_det}) about the equilibrium $a^*$ and $s^*$ by writing  
\begin{eqnarray}
a(t) & = & a^* + \epsilon_a(t) \\
s(t) & = & s^* + \epsilon_s(t),
\end{eqnarray}
where the small perturbations $\epsilon_a$ and $\epsilon_s$ obey the linear equations
\begin{eqnarray}
\frac{ d \epsilon_a}{dt}   & = &  \left ( -\mu + \beta  - \alpha 
-2 \beta a^* - \beta s^*  \right )  \epsilon_a(t)  -  \left (
\alpha  + \beta a^* \right )   \epsilon_s(t)  \label{eps_a} \nonumber \\
& & \\
\frac{ d \epsilon_s}{dt} & = & 
  \mu \left [ \epsilon_a(t) - \epsilon_a(t-\tau)  \right ]. \label{eps_s} 
\end{eqnarray}
The next step is to look for solutions of  the form $\epsilon_a(t) = \epsilon_a(0) e^{\lambda t}$ and $\epsilon_s(t) = \epsilon_s(0) e^{\lambda t}$, where $ \epsilon_a(0)\ll 1$ and $ \epsilon_s(0) \ll 1$ are constants. These solutions exist if  the eigenvalues $\lambda $ satisfy the transcendental equation
\begin{equation}\label{lamb1}
\lambda \left ( \lambda + X \right )  + \left  ( 1-e^{-\lambda \tau} \right ) Y = 0 ,
\end{equation}
where $X = \mu -\beta  + \alpha  +  \beta a^* (2 +\mu \tau) $ and $Y= \mu (\alpha + \beta a^*) $ are auxiliary variables. Setting $\lambda = u + i v$ we get the following equations for the real and imaginary parts of $\lambda$,
\begin{eqnarray}
u^2 -v^2  + u X + Y  & = & Y e^{-u \tau} \cos (v \tau) \label{u1}  \\
-v (2u + X ) & = &  Y e^{-u \tau} \sin (v \tau) \label{v1} .
\end{eqnarray}
 It is clear from these equations that if $v$ is a solution, then so is $-v$, so we can consider $v > 0 $ without loss of generality. 
 Note that $u=v=0$ (i.e., $\lambda = 0$) is a solution. There is another solution with $u=0$ but $v > 0$ which will prove to be important in determining the stability of the fixed point solutions as well as the period of the small amplitude oscillations. By squaring and summing these equations, we can write $v$ in terms of $u$ by solving the quadratic equation
\begin{eqnarray}\label{w}
w^2  +w \left [ 2 u(u+X) +X^2 - 2Y   \right ]  &  & \nonumber \\
+ \left [ u( u+X) +  Y    \right ]^2  -   Y^2 e^{-2u \tau} & = & 0
\end{eqnarray}
with  $w=w(u)= v^2$.  So $u$ is given by  the roots of  $f(x)$, where
\begin{equation}\label{fx}
f(x) =  (2x + X ) \sqrt{w} + Y e^{-x \tau} \sin (\tau \sqrt{w} ) 
\end{equation}
and $w=w(x)$ are the solutions of Eq. (\ref{w}) with $u$ replaced by $x$. Recall that the fixed point solution is unstable if $u= \mbox{Re} (\lambda) > 0$.

\begin{figure}[t] 
\centering
 \includegraphics[width=1\columnwidth]{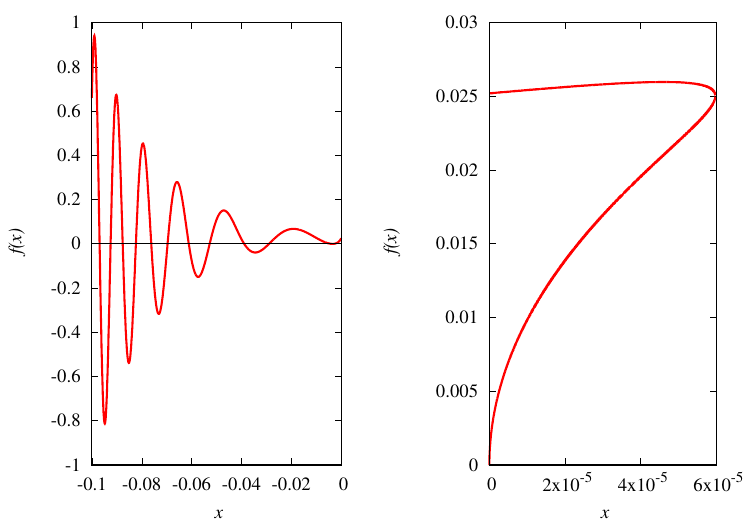}  
\caption{Function $f(x)$ whose roots are the real part of the eigenvalue $\lambda$ for  the rest period $\tau=40$. There are infinitely many negative roots (left panel) but no positive roots  (right panel) so the fixed point $a^* \approx 0.180$, $b^* \approx 0.099$,  $s^* \approx 0.721$ is  stable.
The  parameters are  $\mu = 1/10$, $\alpha = 1/500$ and $\beta=1$.      
 }  
\label{fig:3}  
\end{figure}

  Figure \ref{fig:3} shows the function $f(x)$ defined in Eq,  (\ref{fx}) for $\tau = 40$. There are infinitely many negative roots, but no positive roots, so the fixed point is stable. For $x<0$, the quadratic equation (\ref{w}) has only one positive solution. The other solution is negative and therefore not physical.  For $x>0$, the two solutions of equation (\ref{w}) are positive, so $f(x)$ has two branches. For large $x>0$, the quadratic equation (\ref{w}) has no real roots. From now on we will focus only on the region $x\geq 0$, since a positive root of $f(x)$ would imply the instability of the fixed point.  
  
\begin{figure}[th] 
\centering
 \includegraphics[width=1\columnwidth]{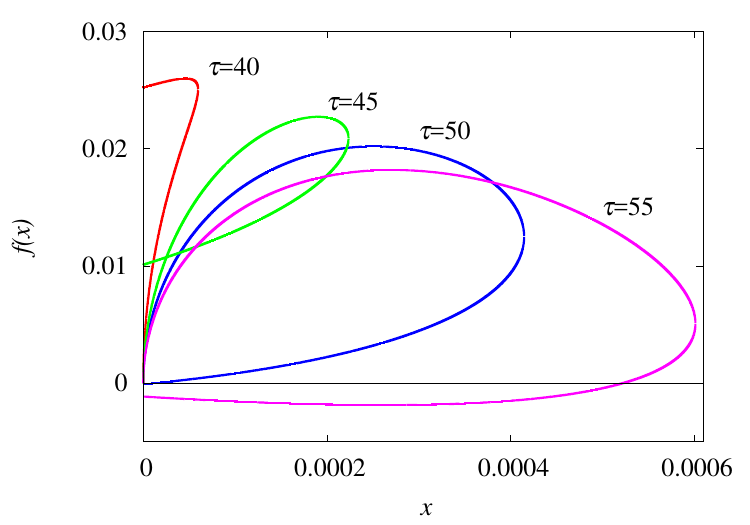}  
\caption{Function $f(x)$ whose roots are the real part of the eigenvalue $\lambda$ for rest period $\tau=40$, $45$, $50$ and $55$ as indicated.
The  parameters are  $\mu = 1/10$, $\alpha = 1/500$ and $\beta=1$.      
 }  
\label{fig:4}  
\end{figure}

  Figure \ref{fig:4} shows $f(x)$ for several values of the rest period $\tau$. The results show that a positive root first appears when $x=0$ is a double root of $f(x)$, i.e. when the two branches of $f(x)$ coincide at $x=0$.  More specifically, if we set $x=0$ in equation (\ref{w})
 we get that the first branch is determined by the solution $w=0$, which gives $ f(0)=0$, and the second branch is determined by $w=2Y-X^2$, which gives $f(0) = g(\tau)$, where
\begin{equation}\label{f0}
g(\tau) = X \sqrt{2Y-X^2}    + Y \sin ( \tau \sqrt{2Y-X^2} ) .
\end{equation}
Imposing $g(\tau) =0$ gives an equation for the value of $\tau=\tau_c$ at which the fixed point solution becomes unstable. This is still quite a formidable equation, since the auxiliary variables $X=X(\tau)$ and $Y=Y(\tau)$ have a complicated dependence on $\tau$ through the fixed point $a^*$. Next, we show how to solve it numerically.

\begin{figure}[th] 
\centering
 \includegraphics[width=1\columnwidth]{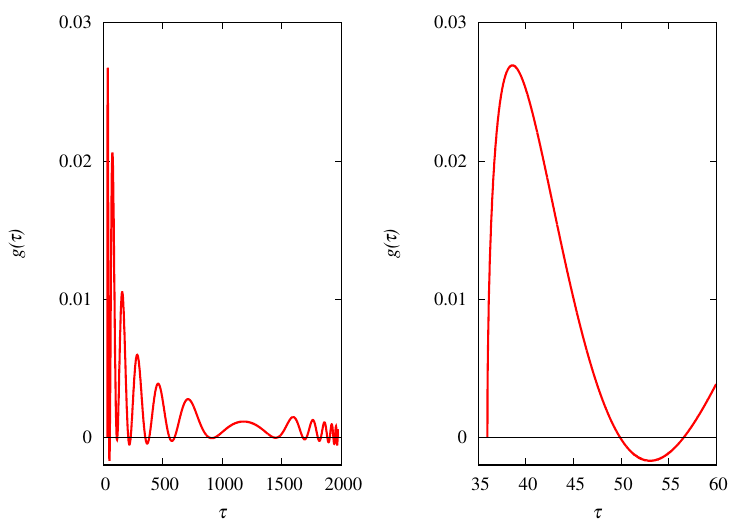}  
\caption{Function $g(\tau)$  whose smallest root  within the open interval $(\tau_l, \tau_u)$ gives the critical rest period $\tau_c$ for $\mu = 1/10$, $\alpha = 1/500$ and $\beta=1$.  The left panel shows $g(\tau)$ in the entire domain $ \tau \in \left [ 35.98,1972.20 \right ]$ and the right panel zooms in on the region around the root $\tau_c \approx 49.902$.
 }  
\label{fig:5}  
\end{figure}

Figure \ref{fig:5} shows the function $g(\tau)$ for the same values of the parameters $\alpha$ and $\mu$ used so far. We recall that $g(\tau)$ gives the values of $f(0)$ obtained with the solution $v^2 =w=2Y-X^2$ shown in Fig.\ \ref{fig:4}. Thus the domain of $g(\tau)$ is determined by the values of $\tau$ for which $w \geq 0$.  We find numerically that, for the parameters used in this figure,  this condition is satisfied in the interval 
$\tau \in \left [ \tau_l, \tau_u \right ]$ where $\tau_l \approx 35.98$  and $\tau_u \approx 1972.20$.  Note that $g(\tau_l)=g(\tau_u)=0$, but the root $\tau= \tau_l$  does not give  $\tau_c$ because $\lim_{x \to 0^+} f(x) > 0$ for $\tau \approx \tau_l$, as shown in Fig.\ \ref{fig:4} and in the right panel of Fig.\ \ref{fig:5}. In words, $f(x)$ does not become negative when $\tau$ deviates from $\tau_l$. In fact, these figures
show that $\tau_c$ is given by the smallest root of $g(\tau)$ within the open interval  $\left (\tau_l,\tau_u \right )$. In particular, we find  $\tau_c \approx 49.902$ (or $\mu \tau_c \approx 4.99$). Thus, the rest period $\tau=50$ used in the original simulations of the autocatalytic ant colony model \cite{Goss_1988} gives truly periodic solutions in the limit of infinitely large populations, but because of its proximity to $\tau_c$ the amplitude of the oscillations is too small to be observed numerically.

\begin{figure}[t] 
\centering
 \includegraphics[width=1\columnwidth]{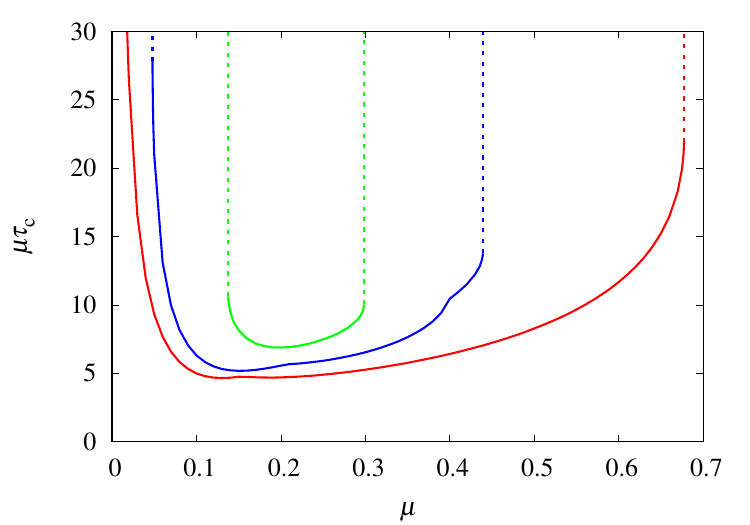}  
\caption{Scaled critical rest period above which the fixed point solutions are unstable as function of the inactivation probability $\mu$ for (bottom to top)  $\alpha = 1/500$, $1/50$ and $1/20$.  As usual, $\beta=1$. The dashed vertical lines indicate the lower $\mu_l$ and upper $\mu_u$ bounds of the region of existence of period solutions for a given $\alpha$. For $\alpha=1/500$ we have $\mu_l \approx 0.0047$. Note the discontinuities in the derivative of $\tau_c$ at some singular points.
 }  
\label{fig:6}  
\end{figure}

Figure \ref{fig:6} shows how the critical rest period $\tau_c$ is affected by the parameters $\mu$ and $\alpha$, which we previously held fixed.
The results show the complexity of the problem. It turns out that the regime of periodic solutions  exists only in a limited region of the parameter space. For a fixed $\alpha$, there is an upper bound $\mu_u$ on the inactivation probability beyond which the fixed-point solution is stable, regardless of how large the rest period $\tau$ is. Note that $\tau_c$ does not diverge at $\mu=\mu_u$: the transition line simply ends at this upper bound because  $g(\tau)$  has no roots except the domain extremes $\tau_l$ and $\tau_u$. The same happens for the lower bound $\mu_l$, below which the fixed point solution is stable.  Most interestingly, increasing $\alpha$ decreases the range of parameters where  the periodic solutions exist, until these solutions disappear altogether at $\alpha = \alpha_c$. In other words, for $\alpha > \alpha_c \approx 0.0589$ the fixed point solutions are stable regardless of the values of $\mu$ and $\tau$.

\begin{figure}[t] 
\centering
 \includegraphics[width=1\columnwidth]{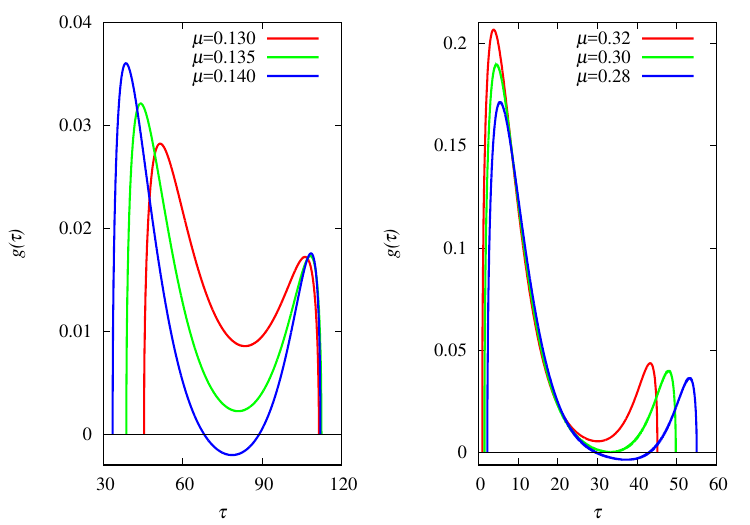}  
\caption{Function $g(\tau)$ whose smallest root  within the open interval $(\tau_l, \tau_u)$ gives the critical rest period $\tau_c$ for  $\alpha = 1/20$ and $\beta=1$.  The left panel shows the results  for  (top to bottom at $\tau=70$) $\mu=0.13$, $0.135$ and $0.14$. The double root  occurs at $\mu = \mu_l \approx 0.1374$ and $\tau_c \approx  79.43$. The right panel 
shows the results for (top to bottom at $\tau=40$) $\mu=0.32$, $0.30$ and $0.28$.   The double root  occurs at $\mu = \mu_u \approx 0.298$ and $\tau_c \approx  33.62$.  
 }  
\label{fig:7}  
\end{figure}

In order to understand the results summarized in Fig.\ \ref{fig:6}, let us look at the case $\alpha = 1/20$ in more detail. Figure \ref{fig:7} shows  what happens for $\mu$ in the vicinity of $\mu_l$ and $\mu_u$ and  indicates how these parameters can be calculated by finding the double root of $g(\tau)$. In fact, near these extremes, this pattern of variation of $g(\tau$) occurs for all values of $\alpha$, so we can use the criterion that $g(\tau)$ becomes negative, leading to the appearance of an eigenvalue with a positive real part, by a double root to determine the region in parameter space where the periodic solutions exist. In practice, this criterion boils down to determining whether the minimum of $g(\tau)$ is positive or negative. We just note that for small $\alpha$, the pattern shown in Fig. \ref{fig:7} appears only very close to $\mu_l$.

\begin{figure}[t] 
\centering
 \includegraphics[width=1\columnwidth]{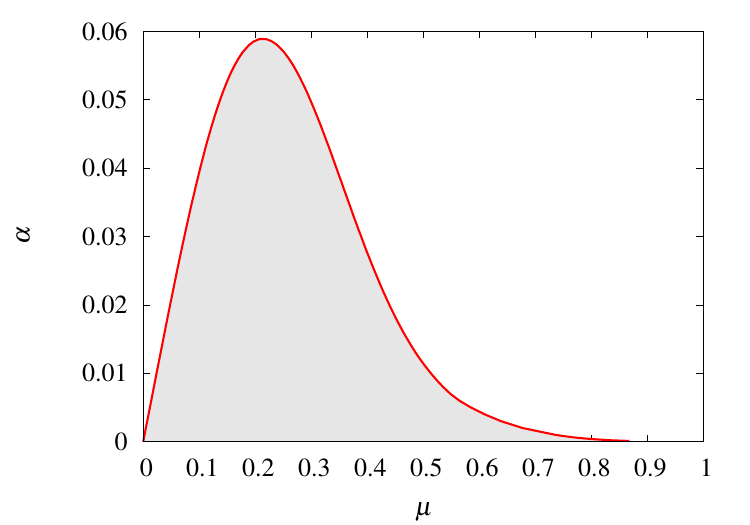}  
\caption{Values of the inactivation probability $\mu$ and the spontaneous activation probability $\alpha$  (shaded region) for which the fixed-point solutions are unstable for some values of the rest period $\tau$ and the delay differential equations exhibit periodic solutions.  As usual, $\beta=1$.
 }  
\label{fig:8}  
\end{figure}

Figure \ref{fig:8} shows  the region in the parameter space $(\mu, \alpha) $ where the fixed-point solutions are unstable  for some  choices of  the rest period  $\tau$. Outside this region, the fixed-point solutions are stable, regardless of the value of $\tau$. The key parameter here is 
the spontaneous activation probability $\alpha$, which acts as a noise that disturbs the synchrony resulting from the autocatalytic activation of the ants by contact between active and activatable inactive ants. As already pointed out, the maximum amount of noise compatible with synchronization is $\alpha = \alpha_c \approx 0.0589$, obtained for $\mu \approx 0.21$. Interestingly,  this disruptive effect of spontaneous activation is enhanced  by either  increasing or decreasing  the probability of inactivation $\mu$, as  already  shown in Fig.\ \ref{fig:6}. Of course, for large $\mu$, an active ant is likely to become inactive before it  finds an activatable inactive ant, reducing the effectiveness of the autocatalytic activation synchronization mechanism. For small $\mu$, there are so many active ants at any time that the synchronizing effect of the  rest period $\tau$ becomes ineffective: whenever an ant wakes up, it is activated.

\begin{figure}[h] 
\centering
 \includegraphics[width=1\columnwidth]{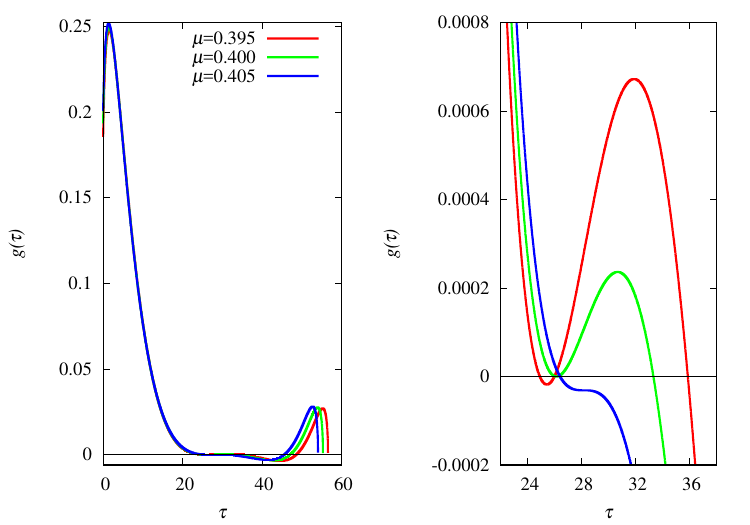}  
\caption{Function $g(\tau)$ whose smallest root  within the open interval $(\tau_l, \tau_u)$ gives the critical rest period $\tau_c$ for  $\alpha = 1/50$ and $\beta=1$.  The left panel shows the results  in the entire domain  for  $\mu=0.395$, $0.400$ and $0.405$ and the right panel zooms in on the vicinity of $\tau_c$.
 }  
\label{fig:9}  
\end{figure}

Another curious feature shown in Fig.\ \ref{fig:6} is the existence of singular points where the derivatives of $\tau_c$ with respect to $\mu$ are discontinuous. The scale of the figure obscures this phenomenon somewhat, but it will become more conspicuous  when we consider the periodic solutions near $\tau_c$. The singularity is due to a change in the character  of the smallest root of $g(\tau)$ as shown in Fig. \ref{fig:9} for $\alpha = 1/50$. What happens is that for some value of $\mu \in \left ( 0.4, 0.405 \right )$, $\tau_c$ is given by a triple root, and merging the three smallest roots (excluding the lower extreme $\tau_l$) of $g(\tau)$ yields a root that behaves differently than the smallest root before the merge. The point here is that these singularities are not  artifacts of the numerical procedure used to find the roots of $g(\tau)$, but, unlike the phenomenon that gives rise to $\mu_l$ and $\mu_u$,    they have  no implications on the nature of the asymptotic solutions of the mean-field equations.

\begin{figure}[h] 
\centering
 \includegraphics[width=1\columnwidth]{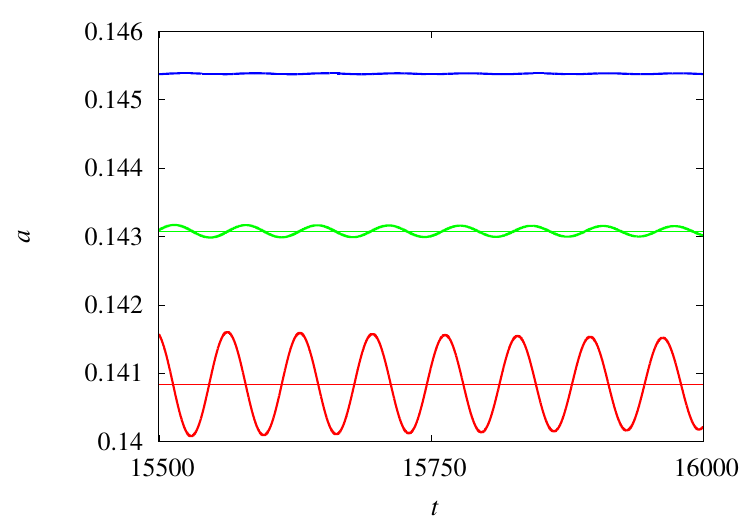}  
\caption{Fraction of active ants $a(t)$ for rest period (top to bottom) $\tau=52, 53$ and $54$.  The other parameters are  $\mu = 1/10$, $\alpha = 1/500$ and $\beta=1$.   The fixed point solutions are stable for $\tau < 49.902$. The thin horizontal  lines indicate the unstable fixed point $a^*$, given by Eq. (\ref{fp_a}). The initial conditions are $a(t)=0$ for $t \in [-\tau,0)$, $a(0)=1$, $b(0)=0$ and $s(0)=0$.  
 }  
\label{fig:10}  
\end{figure}

To sum up, we note that Figs.\ \ref{fig:6} and \ref{fig:8} completely summarize all the results of the fixed point analysis. The other figures are necessary only to explain how  these two key figures are generated.  In particular, Fig.\ \ref{fig:6}  shows that for fixed  $\alpha$  the periodic solutions appear only in the range  $\mu \in [\mu_l,\mu_u ]$ and for $\tau > \tau_c$. Thus, we can see the effects of all three model parameters in this figure (we recall that we can set  $\beta=1$  without loss of generality). Increasing $\alpha$  decreases the range $[\mu_l,\mu_u ]$  where the periodic solutions appear for sufficiently large $\tau$. This result is exhibited in Fig.\  \ref{fig:8}, where the range $[\mu_l,\mu_u ]$  is shaded for each  $\alpha$.  The important point is that outside the shaded region, the fixed point solutions are stable regardless of the value of $\tau$. Since Fig.\ \ref{fig:8} is not drawn for any particular value of $\tau$, it is a strong result: periodic solutions can occur inside the shaded region, provided that $\tau$ is sufficiently large, but they do not occur outside this region.  In addition, apart from the expressions for the fixed-point solutions, Eqs.\ (\ref{fp_b}), (\ref{fp_s}) and (\ref{fp_a}), all our results are numerical, because Eq.\ (\ref{lamb1}) cannot be solved analytically. We recall that the real part of the roots of this equation determines the stability of the fixed-point solutions. In fact, Figs.\ \ref{fig:3}, \ref{fig:4}, \ref{fig:5}, \ref{fig:7}, and \ref{fig:9} illustrate the rather complicated numerical procedure to obtain the real part of the roots of Eq.\  (\ref{lamb1}) and, in particular, to determine the value of the model parameters where roots with positive real part first appear.

\section{Periodic solutions}\label{sec:per}

The first important result about the periodic solutions of the system of delay differential equations (\ref{a_det})-(\ref{s_det}) is that the period $T$ must be different from $\tau$, otherwise $s(t)$ would not be periodic.   The second is that the amplitude of the oscillations becomes arbitrarily small as $\tau$ approaches $\tau_c$ for $\mu$ and $\alpha$ fixed, as shown in Fig.\ \ref{fig:10} for $\mu=1/10$ and $\alpha = 1/500$. The oscillations for $\tau=52$ are barely noticeable on the already narrow scale of the figure, which is why we claim that the large amplitude oscillations observed in the original simulations of the model for $\tau=50$ \cite{Goss_1988}. are due to a finite-size effect, namely, the amplification of  the amplitudes due to coherence resonance \cite{McKane_2005}. However, this result allows to estimate the period $T$ for $\tau \approx \tau_c$, where the model is described by the linear equations (\ref{eps_a}) and (\ref{eps_s}).  In fact, in this case the period is simply $T =2\pi/v$ where $v>0$ is the imaginary part of the eigenvalue $\lambda$ when its real part $u$ vanishes.  We recall that  the condition $u=0$, which implies $v = \sqrt{2Y-X^2}$,  determines the critical rest period $\tau_c$, as shown in  Fig.\ \ref{fig:4}.  We note that the mean-field equations exhibit periodic solutions of large amplitude if we set the model parameters properly, e.g., $\tau \gg \tau_c$.

\begin{figure}[h] 
\centering
 \includegraphics[width=1\columnwidth]{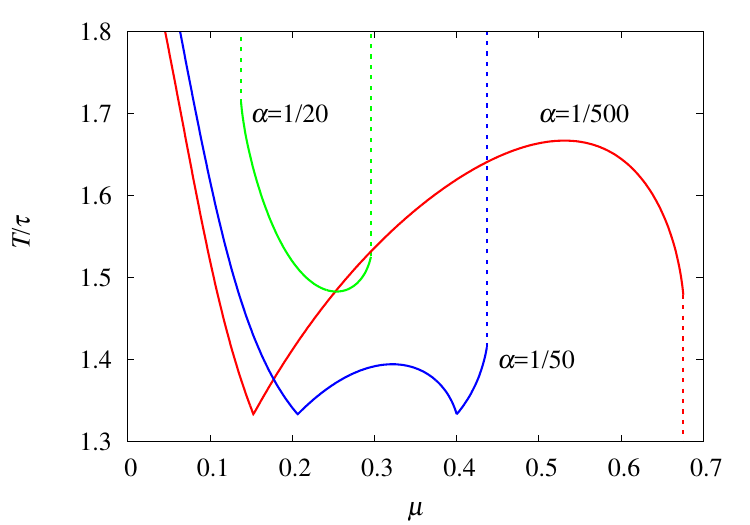}  
\caption{Ratio between the period $T$ of the periodic solutions  and the rest period at $\tau= \tau_c$  for  $\alpha = 1/500$, $1/50$ and $1/20$ as indicated.  As usual, $\beta=1$. The dashed vertical lines indicate the lower $\mu_l$ and upper $\mu_u$ bounds of the region of existence of period solutions for a given $\alpha$.
 }  
\label{fig:11}  
\end{figure}

Figure \ref{fig:11} shows the ratio $T/\tau$ for the periodic solutions near $\tau=\tau_c$.  Recall that the periodic solutions exist only in the interval $\left [ \mu_l, \mu_u \right ]$. Note that both $T$ and $\tau_c$ vary with $\mu$ and $\alpha$, which explains the complexity of the results (the plot of $\mu T$ versus $\mu$ is very similar to Fig.\ \ref{fig:6}), but the rest period is clearly the natural scale for measuring the period of the oscillations. In particular, we find $T > \tau$, as expected. In fact, an ant should go through three stages in a cycle: inactive, which takes exactly a time $\tau$, activatable inactive, which takes  time $1/(\alpha + \beta a^*)$ on  average, and active, which takes time $1/\mu$ on average. So the period  $T$ must be greater than the the rest period $\tau$.   The singularities that were barely noticeable in Fig.\ \ref{fig:6} are now very prominent, but they do not seem to have any physically relevant implications.

%
\section{Discussion}\label{sec:conc} 
%

The general phenomena to which our work is related are synchronized bursts of activity, in which a large fraction of individuals  in a population become active at roughly the same time. These phenomena are of interest when the bursts of activity occur repeatedly as the population evolves over time, producing oscillations in the activity patterns. In fact,  synchronization cannot be seen without oscillations \cite{Strogatz_1993}. The synchronized bursts of activity are not necessarily periodic, and this is the case in the finite-size simulations where the demographic noise produces the coherence resonance phenomenon, which is characterized by oscillations with a narrow frequency distribution and stochastically varying amplitude \cite{McKane_2005}. However, in the absence of noise, the oscillations are periodic, as shown in our mean-field analysis. Synchronized bursts of activity such as  those studied here occur in
 predator-prey models \cite{McKane_2005},  in epidemic models with delays \cite{Hethcote_1981,Sebastian_2011}, and in
neuronal networks \cite{Izhikevich_2000,Majhi_2019}.

The  autocatalytic ant colony model generates periodic synchronized rhythms thanks to the synergy between two key components, viz. the activation of activatable inactive ants by active ants, whose strength is determined by the parameter $\beta$, and the rest period $\tau$ of inactive ants. None of these components alone can produce periodic  solutions to the mean-field equations (\ref{a_det})-(\ref{s_det}) describing an ant colony of infinite size. In fact, on the one hand, Fig.\ \ref{fig:6} shows that there is a minimum value of the rest period $\tau_c > 0$ above which the periodic solutions appear, so that a sufficiently large rest period is a necessary condition for observing periodic synchronized rhythms in the model.  On the other hand, Fig.\ \ref{fig:8} shows that for small values of $\beta$, which correspond to large values of $\alpha$, there are no periodic solutions to the mean-field equations, regardless of the value of the rest period $\tau$. As a matter of fact, this figure shows that the range of parameters for which the periodic solutions exist is very limited.  We note that  in the original simulations of the model using a colony of size $N=50$, unsynchronized activity was reported when the rest period $\tau$ was too short relative to the activity period ($1/\mu$) or when the activation coefficient $\beta$ was too low \cite{Goss_1988}.

We emphasize that our only analytical evidence for considering the periodic solutions as stable limit cycles, i.e., when perturbed, the solution returns to the original periodic solution in the limit $t \to \infty$ and the periodic behavior is independent of the initial conditions, is the instability of the fixed-point solutions (the real part of the eigenvalues $\lambda$ given by Eq.\ (\ref{lamb1}) are positive).   This contrasts with the periodic solutions of conservative systems, for which the real part of the eigenvalues of the community matrix are zero and the phase space trajectories are closed orbits depending on the initial conditions \cite{Murray_2007}.  Since this evidence is certainly not conclusive, we complemented it with an extensive numerical study of the mean-field equations using different initial conditions for the model parameters in the regions where the fixed-point solutions are unstable. The results  showed that the periodic solutions are indeed the only attractors of the dynamics in these regions of the model parameters.

The neatness of the autocatalytic ant colony model \cite{Goss_1988} makes it easy to implement through Monte Carlo simulations, but the results are not easy to interpret without knowledge of the nature of the equilibrium solutions. Our main contribution is the derivation of mean-field delay differential equations, which are exact in the limit $N \to \infty$, but give a good fit to the simulation results already for colony sizes on the order of a few thousand ants, as shown in Fig.\ \ref{fig:2}.  A standard, but rather complicated, analysis of the equilibrium solutions of these equations indicates that both the unsynchronized and synchronized activity patterns observed in the short-time simulations for small $N$ are fluctuations due to finite-size effects. In fact, these fluctuations make it virtually impossible to distinguish true periodic solutions from damped oscillations that eventually settle to a fixed point, as shown in Figs.\ \ref{fig:1} and \ref{fig:2}.
In particular, the reported unsynchronized activity patterns \cite{Goss_1988} are fluctuations around stable fixed points.  Note, however, that the colony sizes used in the laboratory experiments range from $7$ to $250$ ants, so the small $N$ simulations are probably more realistic than our mean-field formulation. The large finite-size fluctuations make it very difficult to elucidate the role of the model parameters and, in particular, to appreciate the various threshold phenomena associated with the instability of the fixed-point solutions.

The uniform rest period assumption of the original autocatalytic ant colony model \cite{Goss_1988} is a great simplification, but relaxing it by assuming, for example, that the rest periods are exponentially distributed random variables \cite{Doering_2022} would lead to very different mean field equations: Instead of delay differential equations, the model with random rest periods would be described by integro-differential equations (see \cite{Hethcote_1981} for an analytical study of an epidemic model with random rest periods).  Note that finite population simulations may not be very helpful, since the effect of heterogeneity in the rest periods is likely to be masked by the stochastic resonance effects due to demographic noise. Relaxing this assumption  is a challenging step in the analytical exploration of the autocatalytic ant colony model.

Finally, we note that there is no consensus biological explanation for the existence of short-term activity cycles in  colonies of some ant species.  Although there are suggestions that synchronized activity promotes more efficient brood care \cite{Hatcher_1992} and empirical evidence that it can improve physical access to all portions of the nest as inactive ants can act as immobile obstacles to moving ants \cite{Doering_2023}, another  explanation is that it is an inevitable result of interactions between ants, without any adaptive significance, i.e., it is an epiphenomenon \cite{Cole_1991}. For colonies of infinite size, our analysis shows that the periodic solutions are far from inevitable in the sense that the fixed-point solutions are stable in large regions of the parameter space. In particular,  periodic solutions exist only for   $\alpha < 0.06$ (see Fig. \ref{fig:8}), where $\alpha$ is the spontaneous activation probability.  Of course, the model parameters may be subject to selective pressures, an issue that can be addressed using a group selection approach, similar to what has been done to evolve response thresholds \cite{Bonabeau_1998} in the context of  division of labor \cite{Duarte_2012,Fontanari_2024}. However, in a more realistic scenario where the  model parameters are stochastic rather than fixed variables, the oscillatory dynamic  behavior is likely to be more common. Moreover, for finite populations, the coherence resonance phenomenon will sustain oscillations for very large $N$ \cite{McKane_2005} even in the parameter regions where the fixed-point solutions are stable. These noise effects are likely present in the synchronized bursts of activity that have  been reported for fire ants, which have large colony sizes consisting of thousands of ants \cite{Tennenbaum_2017}.  Therefore, even from a theoretical perspective, activity cycles are likely to be more common than predicted by our mean-field model, which can nevertheless be a starting point for further improvements.

\bigskip

\acknowledgments
JFF is partially supported by  Conselho Nacional de Desenvolvimento Ci\-en\-t\'{\i}\-fi\-co e Tecnol\'ogico -- Brasil (CNPq) -- grant number 305620/2021-5.  PMMS is supported by  Coordena\c{c}\~ao de Aperfei\c{c}oamento de Pessoal de N\'{\i}vel Superior -- Brasil (CAPES) -- Finance Code 001.

\end{document}